\documentclass[11pt,a4paper]{article}
\usepackage[hyperref]{naaclhlt2019}
\usepackage{times}
\usepackage{latexsym}
\usepackage{algorithm}
\usepackage{times}
\usepackage{latexsym}
\usepackage{times}  
\usepackage{helvet}  
\usepackage{courier}  
\usepackage{graphicx}  
\usepackage{algorithm}
\usepackage{booktabs}
\usepackage{graphicx}
\usepackage{amsmath}
\usepackage{comment}
\usepackage{amsfonts}
\usepackage{xcolor}
\usepackage{multirow}
\usepackage{verbatim}
\usepackage{multicol}
\usepackage{subfigure}

\aclfinalcopy 
\def\aclpaperid{147} 

\title{Long-tail Relation Extraction via  Knowledge Graph Embeddings and Graph  Convolution  Networks}

\author{Ningyu Zhang\textsuperscript{1,2,3}   \quad
Shumin Deng\textsuperscript{1,3}    \quad
 Zhanlin Sun\textsuperscript{1,3}  \quad
  Guanying Wang\textsuperscript{1,3}  \\
 \textbf{Xi Chen\textsuperscript{4}}  \quad
  \textbf{Wei Zhang\textsuperscript{2,3}} \quad
   \textbf{Huajun Chen\textsuperscript{1,3}}\thanks{\quad Corresponding author.} \\
 1. College of Computer Science and Technology, Zhejiang University\\
 2. Alibaba Group \\
 3. AZFT\thanks{Alibaba-Zhejiang University Frontier Technology Research Center}  Joint Lab for Knowledge Engine\\
 4. Tencent\\
  \{3150105645,231sm,guanying\_wgy,huajunsir\}@zju.edu.cn \\
  jasonxchen@tencent.com,\{ningyu.zny,lantu.zw\}@alibaba-inc.com}
  
\date{}

\begin{document}
\maketitle
\begin{abstract}
We propose a distance  supervised  relation extraction approach  for long-tailed, imbalanced data which is prevalent in real-world settings. Here, the challenge is to learn accurate "few-shot" models for classes existing at the tail of the class distribution, for which little data is available. Inspired by the rich semantic correlations between classes at the long tail and those at the head, we take advantage of the knowledge  from data-rich classes at the head of the distribution to boost the performance of the data-poor classes at the tail.   First, we propose to leverage implicit  relational knowledge among class labels  from knowledge graph embeddings and  learn explicit   relational knowledge using graph  convolution  networks. Second, we integrate that relational knowledge into relation extraction model by coarse-to-fine knowledge-aware attention mechanism.  We demonstrate our results for a large-scale benchmark dataset which show that our approach significantly outperforms other baselines, especially for long-tail relations.
\end{abstract}

\section{Introduction}
Relation extraction (RE) is an important task in information extraction, aiming to extract the relation between two given entities based on their related context. Due to the capability of extracting textual information and benefiting many NLP applications (e.g., information retrieval, dialog generation, and question answering), RE appeals to many researchers. Conventional supervised models have been widely explored in this task \cite{zelenko2003kernel,zeng2014relation}; however, their performance heavily depends on the scale and quality of training data. 
\begin{figure}
\centering
\includegraphics [width=0.4\textwidth]{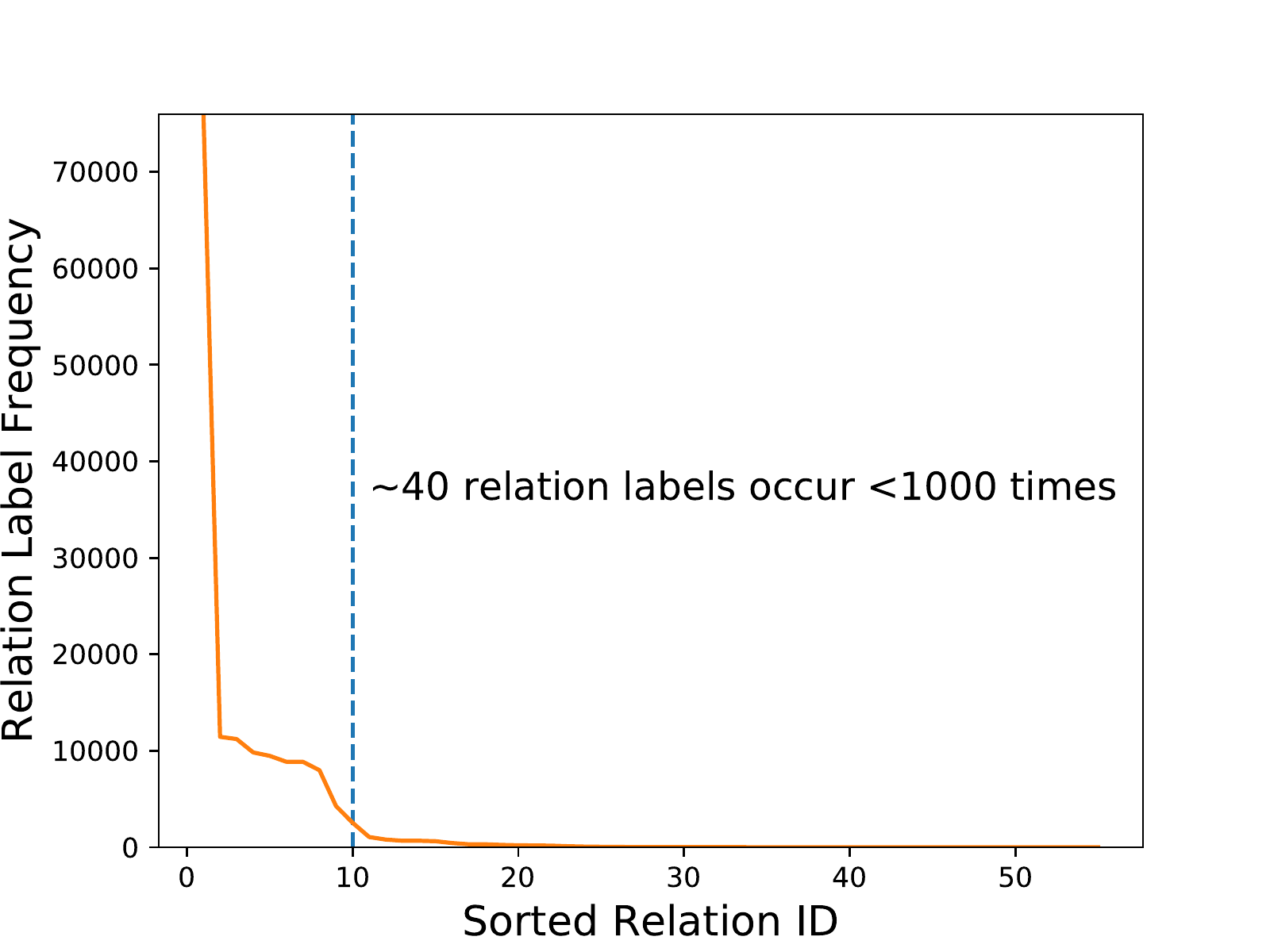}
\caption{Label frequency distribution of classes without $NA$  in NYT dataset.}
\label{pic1}
\end{figure}

To construct large-scale data, \cite{mintz2009distant} proposed a novel distant supervision (DS) mechanism to automatically label training instances by aligning existing knowledge graphs (KGs) with text. DS enables RE models to work on large-scale training corpora and has thus become a primary approach for RE recently \cite{wu2017adversarial,feng2018reinforcement}.  
Although these DS models achieve promising results on common relations, their  performance still degrades dramatically when there are only a few training instances for some relations. Empirically,  DS can automatically annotate adequate amounts of training data; however, this data usually only covers a limited part of the relations. Many relations are long-tail and still suffer from data deficiency. Current DS models ignore the problem of long-tail relations, which makes it challenging to extract comprehensive information from plain text. 

Long-tail relations are important and cannot be ignored. Nearly  70\% of the relations  are long-tail in the widely used New York Times (NYT) dataset\footnote{http://iesl.cs.umass.edu/riedel/ecml/} \cite{riedel2010modeling,lei2018cooperative} as shown in Figure \ref{pic1}. 
Therefore, it is crucial for models to be able to extract relations with limited numbers of training instances.

\begin{figure}
\centering
\includegraphics [width=0.5\textwidth]{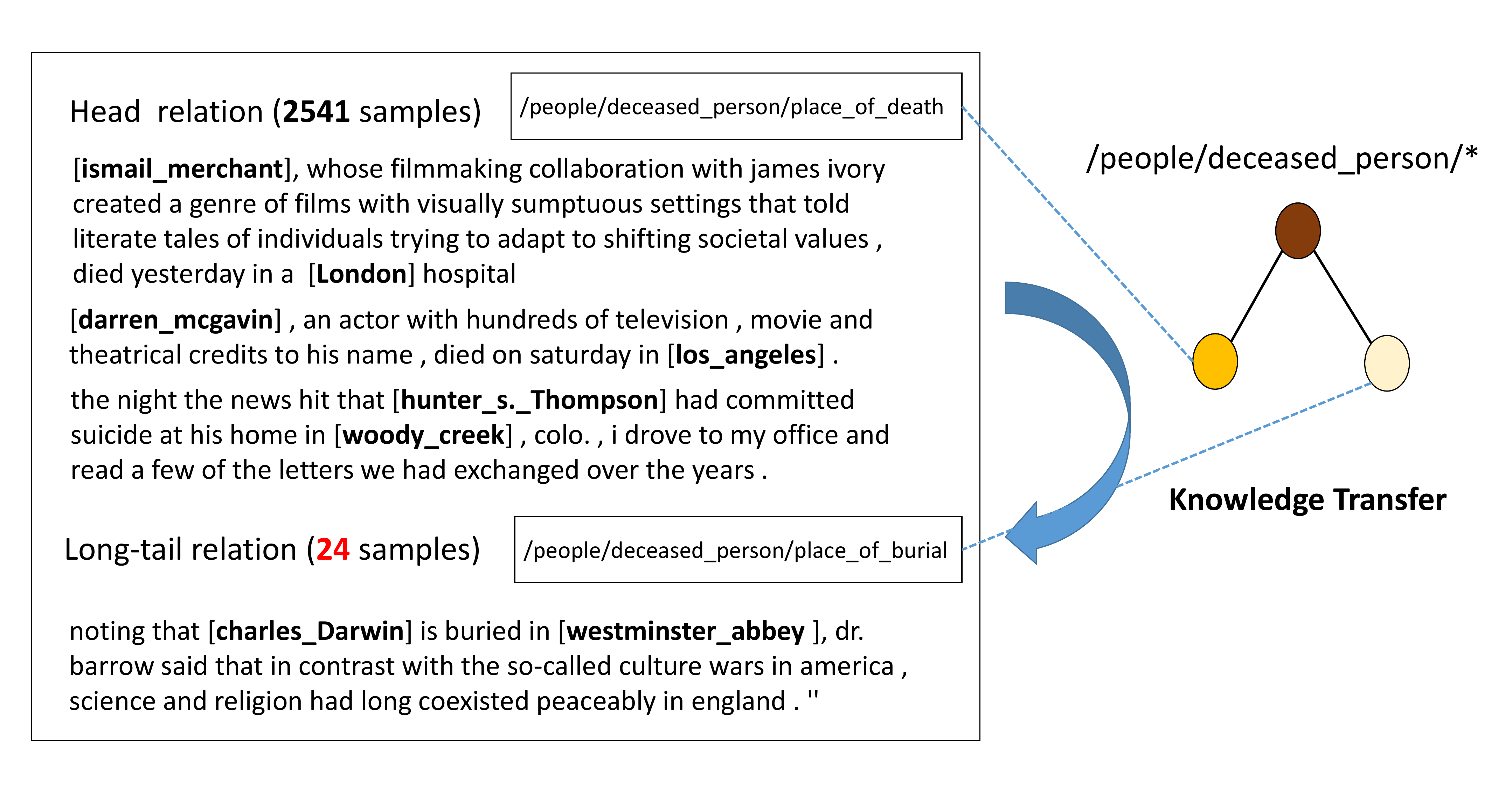}
\caption{Head and long-tail relations.}
\label{pic2}
\end{figure}

Dealing with long tails is very  difficult as few training examples are available. Therefore, it is natural  to transfer knowledge from data-rich and  semantically similar head  classes to data-poor tail classes \cite{wang2017learning}. For example,   the  long-tail relation  /people/deceased\_person/place\_of\_burial and head relation /people/deceased\_person/place\_of\_death are in the same branch /people/deceased\_person/*  as shown in  Figure \ref{pic2}. They are semantically similar, and  it is beneficial to leverage head relational  knowledge and transfer it to the long-tail relation, thus enhancing general performance.  In other words,  long-tail relations of one entity tuple can have class ties with head relations, which can be leveraged to enhance RE for narrowing potential search spaces and reducing uncertainties between relations when predicting unknown relations \cite{ye2017jointly}. If one pair of entities contains /people/deceased\_person/place\_of\_death, there is a high probability that it will contain /people/deceased\_person/place\_of\_burial. If we can incorporate the relational knowledge  between two relations, extracting head relations  will provide evidence for the prediction of long-tail relations.

However,  there exist two problems: (1) \emph{Learning relational knowledge:} Semantically similar classes  may contain  more relational  information that will boost transfer, whereas   irrelevant classes (e.g., /location/location/contains and /people/family/country) usually contain less relational  information that may result in  negative transfer.  (2) \emph{Leveraging relational  knowledge:}  Integrating relational knowledge to existing RE models is challenging. 

To address the problem of  learning relational knowledge,  as shown in \cite{lin2016neural,ye2017jointly}, we use class embeddings to represent relation classes and  utilize KG embeddings and graph convolution networks (GCNs) to extract implicit  and  explicit  relational  knowledge.  Specifically, previous studies \cite{yang2014embedding} have shown that the embeddings of semantically similar relations are located near each other in the latent space.  For instance, the relation /people/person/place\_lived and   /people/person/nationality  are more relevant,   whereas  the relation  /people/person/profession  has  less correlation with the former two relations. Thus,  it is  natural to leverage  this   knowledge from KGs. However, because there are many one-to-multiple relations in KGs, the relevant information for each  class may be scattered. In other words, there may not be enough relational signal between classes.  Therefore, we utilize GCNs  to learn explicit  relational knowledge. 

To address the problem of leveraging relational knowledge, we first use convolution neural  networks \cite{zeng2014relation,zeng2015distant} to encode sentences; then introduce coarse-to-fine  knowledge-aware attention mechanism for  combining relational knowledge with encoded sentences into bag representation vectors.  The relational  knowledge not only provides more information for relation prediction but also provides a better reference message for the attention module to raise the performance of long-tail classes.

Our experimental results on the NYT dataset show  that: (1) our model is effective compared to baselines especially for  long-tail relations; (2) leveraging relational knowledge  enhances RE and our model is efficient in learning relational knowledge via   GCNs.




\section{Related Work} 
\textbf{Relation Extraction.} Supervised RE models  \cite{zelenko2003kernel,guodong2005exploring,mooney2006subsequence} require adequate amounts of annotated data for training which is  time-consuming. Hence, \cite{mintz2009distant} proposd DS to automatically label data. DS inevitably accompanies with the wrong labeling problem. To alleviate the noise issue, \cite{riedel2010modeling,hoffmann2011knowledge} proposed multi-instance learning (MIL) mechanisms.  Recently, neural models have been widely used for RE;  those models can accurately capture textual relations without explicit linguistic   analysis \cite{zeng2015distant,lin2016neural,zhang2018capsule}. To further improve the  performance, some studies incorporate external information \cite{zeng2016incorporating,ji2017distant,han2018neural} and advanced training strategies \cite{ye2017jointly,liu2017soft,huang2017deep,feng2018reinforcement,zeng2018large,wu2017adversarial,qin2018dsgan}.
These works mainly adopt DS to make large-scale datasets and reduce the noise caused by DS, regardless of the effect of long-tail relations.
 
There are only a few studies on long-tail for RE \cite{gui2016exploring,lei2018cooperative,han2018hierarchical}. Of these, \cite{gui2016exploring}  proposed an explanation-based  approach,  whereas \cite{lei2018cooperative} utilized external knowledge (logic rules). These studies treat  each relation in isolation, regardless of the rich semantic correlations between the relations.   \cite{han2018hierarchical}  proposed a  hierarchical attention scheme for RE, especially  for long-tail relations.  Different from those approaches,  we leverage  implicit  and explicit relational  knowledge from KGs  and GCNs rather than data-driven learned parameter spaces where similar relations may have distinct parameters, hindering the  generalization of long-tail classes.

\textbf{Knowledge Graph Embedding.} 
Recently, several KG embedding  models have been proposed. These methods learn low-dimensional vector representations for entities and relations \cite{bordes2013translating,wang2014knowledge,lin2015learning}. TransE \cite{bordes2013translating} is one of the most widely used models, which views relations as translations from a head entity to a tail entity on the same low-dimensional hyperplane.    Inspired by  the rich knowledge in KGs,   recent works \cite{han2018neural,wang2018label,lei2018cooperative} extend   DS models under the guidance of KGs. However, these works  neglect rich correlations between relations.  Relation structure (relational knowledge) has been studied and  
is quite effective for KG completion \cite{zhang2018knowledge}. 
To the best of our knowledge, this is the first effort to consider the  relational knowledge of classes (relations) using KGs  for RE.

\textbf{Graph Convolutional Networks.} 
GCNs generalize CNNs beyond two-dimensional and one-dimensional spaces. \cite{defferrard2016convolutional} developed spectral methods to perform efficient graph convolutions. \cite{kipf2016semi} assumed the graph structure is known over input instances and apply GCNs for semi-supervised learning. GCNs were applied to relational data (e.g., link prediction) by \cite{schlichtkrull2018modeling}. GCNs have also had success in other NLP tasks such as semantic role labeling \cite{marcheggiani2017encoding}, dependency parsing \cite{strubell2017dependency}, and machine translation \cite{bastings2017graph}. 

Two GCNs studies share similarities with our work. (1)  \cite{chen2017graph} used GCNs on structured label spaces. However, their experiments do not handle long-tail labels and  do not incorporate attention but use an average of word vectors to represent each document. (2) \cite{rios2018few}   proposed a   few-shot and  zero-shot  text  classification method by exploiting structured label spaces with GCNs.  However, they used  GCNs in the label graph   whereas we  utilize  GCNs in the  hierarchy graph of labels.

\section{Methodology}

In this section, we introduce the overall framework of our approach for RE, starting with the notations.
\subsection{Notations}
 
We denote a KG as $\mathcal{G}$ = $\mathcal{E},\mathcal{R},\mathcal{F}$, where $\mathcal{E}$, $\mathcal{R}$ and $\mathcal{F}$ indicate the sets of entities, relations and facts respectively. $(h,r,t) \in \mathcal{F} $ indicates that there is a relation $r \in  \mathcal{R}$ between $h \in \mathcal{E}$  and $t \in \mathcal{E}$. We follow the MIL setting and split all instances into multiple entity-pair bags
$\{\mathcal{S}_{h_1,t_1},\mathcal{S}_{h_2,t_2},...\}$. Each bag $\mathcal{S}_{h_i,t_i}$  contains multiple
instances $\{s_1,s_2,...\}$ mentioning both  entities $h_i$ and $t_i$.  Each instance $s$ in these bags is denoted as a word sequence $s = \{w_1,w_2,...\}$.

\subsection{Framework}
\begin{figure*} \centering

  \includegraphics[width=0.6\textwidth]{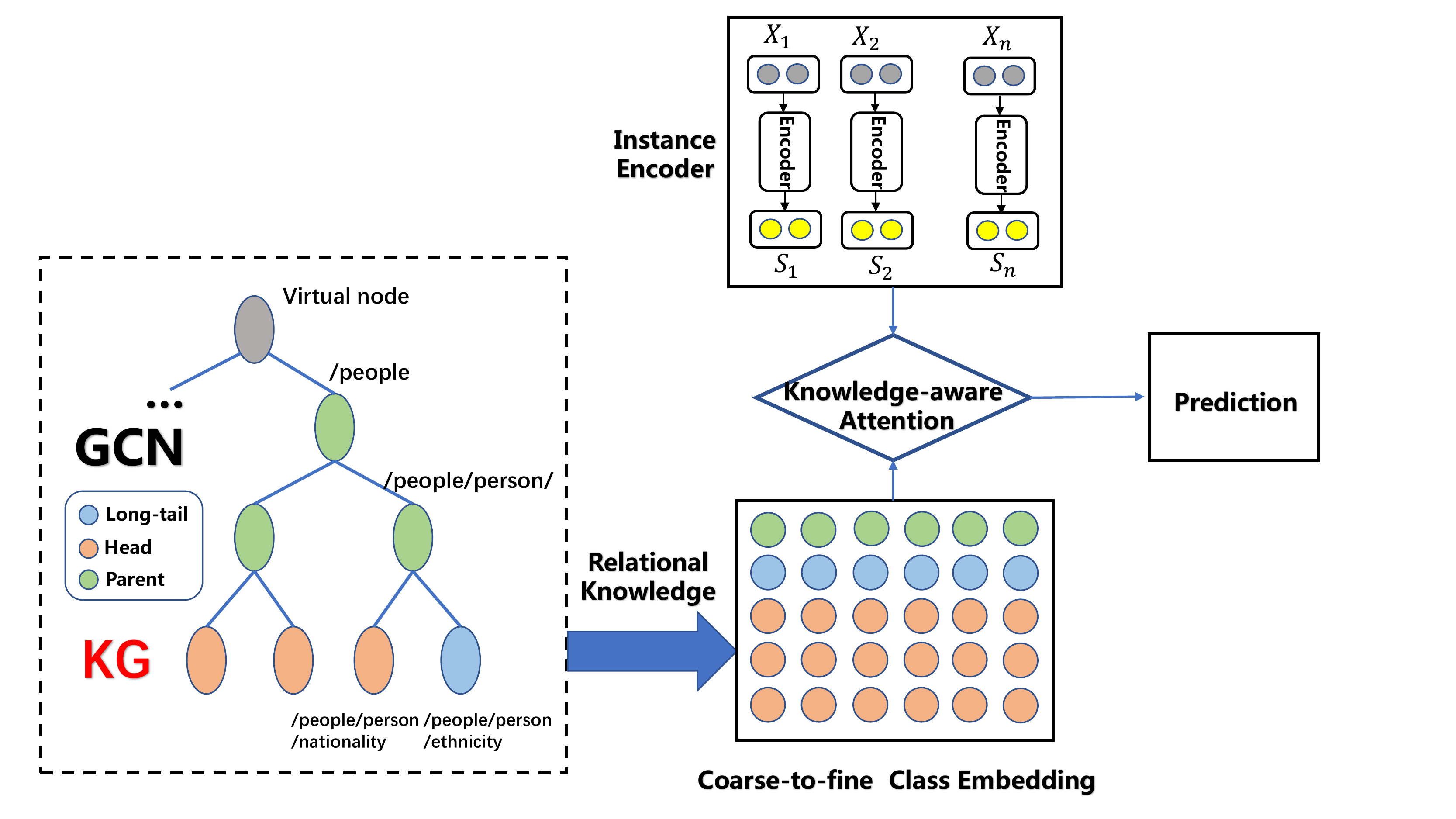}

\caption{Architecture of our proposed model.}
\label{framework}
\end{figure*}

 Our model consists of three parts as shown in  Figure \ref{framework}:

\textbf{Instance Encoder.} Given an instance and its mentioned entity pair, we employ neural networks to encode the instance semantics into a vector. In this study, we implement the instance encoder with convolutional neural networks (CNNs) given both model performance and time efficiency. 

\textbf{Relational Knowledge Learning.} 
Given pretrained KG embeddings (e.g., TransE \cite{bordes2013translating}) as implicit relational knowledge, we employ  GCNs to learn explicit relational knowledge. By assimilating generic message-passing inference algorithms with  neural-network counterpart, we can learn  better embeddings for Knowledge Relation.  We concatenate the outputs of GCNs and the  pretrained KG embeddings  to form the final class embeddings. 

\textbf{Knowledge-aware Attention.}   Under the guidance of final class embeddings, knowledge-aware attention is aimed to select the most informative instance exactly matching relevant relation.

\subsection{Instance Encoder}

Given an instance $s = \{w_1,...,w_n\}$ mentioning two entities, we  encode the raw instance into a continuous low-dimensional vector $x$, which consists of an embedding layer and an encoding layer.

\textbf{Embedding Layer.}
The embedding layer is used to map discrete words in the instance into continuous input embeddings. Given an instance $s$, we map each word $w_i$ in the instance to a real-valued pretrained Skip-Gram \cite{mikolov2013efficient} embedding $w_i \in \mathbb{R}^{d_w}$ .  We adopt position embeddings following \cite{zeng2014relation}. For each word $w_i$, we embed its relative distances to the two entities into two $d_p$-dimensional vectors. We then concatenate the word embeddings and position embeddings to achieve the final input embeddings for each word  and  gather all the input embeddings in the instance.  We thus obtain an embedding sequence ready for the encoding layer.

\textbf{Encoding Layer.}
The encoding layer aims to compose the input embeddings of a given instance into its corresponding instance embedding. In this study, we choose two convolutional neural architectures, CNN \cite{zeng2014relation} and PCNN \cite{zeng2015distant} to encode input embeddings into instance embeddings. Other neural architectures such as recurrent neural networks  \cite{zhang2015relation} can also be used as sentence encoders. Because previous works show that both convolutional and recurrent architectures can achieve comparable state-of-the-art performance, we select convolutional architectures in this study. Note that, our model  is  independent of the encoder choices, and  can, therefore, be easily adapted to fit other encoder architectures.

\subsection{Relational Knowledge Learning through KG Embeddings and GCNs.}
Given  pretrained KG embeddings and predefined class (relation) hierarchies\footnote{For datasets without predefined relation hierarchies, hierarchy clustering \cite{johnson1967hierarchical} or  K-means  can  construct relation hierarchies \cite{zhang2018knowledge}; details can be found in  supplementary materials.},  we first leverage the implicit relational knowledge from KGs and  initialize the hierarchy  label graph;  then we apply  two layer GCNs to  learn  explicit fine-grained relational knowledge from the label space.

\textbf{Hierarchy  Label  Graph  Construction.}
Given a relation set $\mathcal{R}$ of a KG $\mathcal{G}$ (e.g., Freebase), which consists of base-level relations (e.g., /people/person/ethnicity), we can generate the corresponding higher-level relation set $\mathcal{R}^H$. The relations in a high-level set (e.g., people) are more general and common; they  usually contain several sub-relations in the base-level set. The relation hierarchies are tree-structured, and the generation process can be done recursively.  We use a virtual father node to construct  the highest level  associations between relations as shown in Figure \ref{framework}.
In practice, we start from $\mathcal{R}^0 = R$ which is the set of all relations we focus on for RE, and the generation process is performed $L-1$ times to get the hierarchical relation sets $\{\mathcal{R}^0,\mathcal{R}^1,...,\mathcal{R}^{L}\}$, where $\mathcal{R}^{L}$ is the  virtual father node. Each node has a specific type $t \in \{0,1,...,L\}$ to identify its layer hierarchies.  For example, as shown in  Figure \ref{framework},  node /people/person/ethnicity  has a specific type $0$ to indicate it is in the bottom layer of the  graph.  The vectors of each node  in the bottom layer are initialized  through pretrained TransE \cite{bordes2013translating} KG embeddings. Other KG embeddings such as TransR \cite{lin2015learning} can also be adopted. Their parent nodes are initialized by averaging all  children vectors. For example, the node vector of /people/person/ is initialized by averaging all the nodes under the branch of /people/person/* (all child nodes).

\textbf{GCN Output Layer.}  
Due to  one-to-multiple relations and incompleteness in KGs, the implicit relevant information obtained by KG embeddings for each  label is not enough. Therefore, we apply GCNs to learn  explicit relational knowledge among labels.  We  take advantage of the structured knowledge  over our label space using a two-layer GCNs.  Starting with the pretrained relation embedding $v_i^{implicit} \in \mathbb{R}^{d}$  from KGs, we combine the label vectors of the children and parents for the $i$-th label to form,

\begin{equation}
    v_i^1 = f(W^1v_i+\sum\limits_{j \in \mathcal{N}_p} \frac{W_p^1v_j}{|\mathcal{N}_p|} +\sum\limits_{j \in \mathcal{N}_c} \frac{W_c^1v_j}{|\mathcal{N}_c|}  +b_g^1 )
\end{equation}

where $W^1 \in \mathbb{R}^{q*d}$,  $W^1_p \in \mathbb{R}^{q*d}$, $W^1_c \in \mathbb{R}^{q*d}$, $b_g^1 \in \mathbb{R}^q$, $f$ is the rectified linear unit \cite{nair2010rectified} function, and $\mathcal{N}_c$ ($\mathcal{N}_p$) is the index set of the $i$-th label’s children (parents). We use different parameters to distinguish each edge type where parent edges represent all   edges  from high level labels and child edges represent all edges from low level  labels. The second layer follows the same formulation as the first layer and outputs $v_i^{explicit}$. Finally, we concatenate both pretrained $v_i^{implicit}$ with GCNs node vector $v_i^{explicit}$ to form  hierarchy class embeddings, 

\begin{equation}
    q_r=v_i^{implicit} || v_i^{explicit}
\end{equation}

where $q_r \in \mathbb{R}^{d+q}$.

\subsection{Knowledge-aware Attention}
Traditionally, the output layer of PCNN/CNN would learn label specific parameters optimized by a  cross-entropy loss. However, the label specific parameters spaces are unique to each relation, matrices associated with the long-tails  can only be exposed to very few facts during training, resulting in poor generalization.  Instead, our method  attempts to match sentence vectors to their corresponding class embeddings rather than learning label specific attention parameters. In essence, this becomes a retrieval problem. Relevant information from class embeddings contains useful relational knowledge for long-tails among labels.

Practically, given the entity pair $(h, t)$ and its bag of instances $S_{h,t} = \{s_1,s_2,...,s_m\}$, we achieve the instance embeddings $\{s_1,s_2,...,s_m\}$ using the sentence encoder.  We group the  class embeddings according to their type (i.e., according to their layers  in the hierarchy  label graph), e.g.,  $q_{r^i},  i \in \{0,1,...,L \}$. We adopt  $q_{r^i}, i\neq L$ (layer $L$ is the virtual father node) as layer-wise attention query vector. Then, we apply  coarse-to-fine knowledge-aware  attention to them to obtain the textual relation representation $r_{h,t}$. For a relation $r$, we construct its hierarchical chain of parent relations $(r^0,...,r^{L-1})$ using the hierarchy  label graph, where $r^{i-1}$ is the sub-relation of $r^i$. We  propose the following formulas  to compute the attention weight (similarity or relatedness) between each instance’s feature vector $s_k$ and $q_{r^i}$,

\begin{equation}
\begin{aligned}
  e_k &= W_s(tanh[s_k;q_{r^i}])+b_s\\
 \alpha_k^i &=    \frac{exp(e_k)}{\sum_{j=1}^{m}exp(e_j)}
\end{aligned}\label{attention}
\end{equation}
where $[x_1; x_2]$ denotes the vertical concatenation of $x_1$ and $x_2$, $W_s$  is  the weight matrix,  and $b_s$ is the bias. We compute attention operations on  each layer of hierarchy  label graph to obtain corresponding textual relation representations, 
\begin{equation}
r_{h,t}^i = ATT(q_{r^i},\{s_1,s_2,...,s_m\})
\end{equation}

Then we need to combine the relation representations on different layers. Direct concatenation of all the representations is a   straightforward choice.  However,  different layers have different contributions for different tuples. For example, the  relation  /location/br\_state/   has only one sub-relation  /location/br\_state/capital, which indicates that it is more important. In other words, if the sentence has high attention  weights on relation   /location/br\_state/, it has a very high probability to have relation /location/br\_state/capital.  Hence, we use an  attention mechanism to emphasize the layers,  
\begin{equation}
\begin{aligned}
 g_i &= W_g tanh(r_{h,t})\\
    \beta_i&=\frac{exp(g_i)}{\sum_{j=0}^{L-1} exp(g_j)}
    \end{aligned}
\end{equation}

where $W_g$ is a weight  matrix,  $r_{h,t}$ is referred to as a query-based function that scores how well the input textual relation representations and the predict relation $r$ match. The textual relation representations in each layer are computed as, 
\begin{equation}
   r_{h,t}^i = \beta_i r_{h,t}^i
\end{equation}
We simply concatenate the  textual relation representations on different layers as the final representation, 
\begin{equation}
r_{h,r}=Concat(r_{h,t}^0,..,r_{h,t}^{L-1})
\end{equation}
The representation $r_{h,t}$ will be finally fed to compute
the conditional probability $\mathcal{P}(r|h,t,\mathcal{S}_{h,t})$,
\begin{equation}
    \mathcal{P}(r|h,t,\mathcal{S}_{h,t}) = \frac{exp(o_r)}{\sum_{\tilde{r} \in R}exp(o_{\tilde{r}})}
    \end{equation}
where $o$ is the score of all relations defined as,
\begin{equation}
    o=M r_{h,t}
\end{equation}
where $M$ is the representation matrix to calculate the relation scores.  Note that,  attention weight $q_{r^i}$ is  obtained from the outputs of GCNs and pretrained KG embeddings,  which can  provide more informative parameters than data-driven learned parameters, especially for long-tails.  

\section{Experiments}
\subsection{Datasets and Evaluation}
We evaluate our models on the NYT dataset developed by \cite{riedel2010modeling}, which has been widely used in recent studies \cite{lin2016neural,liu2017soft,wu2017adversarial,feng2018reinforcement}. The dataset has 53 relations including the $NA$ relation,  which indicates that the relations of instances are not available. The training set has 522611 sentences, 281270 entity pairs,  and 18252 relational facts. In the test set, there are 172448 sentences, 96678 entity pairs, and 1950 relational facts. In both  training and test set, we truncate sentences with more than 120 words into 120 words.

We evaluate all models in the held-out evaluation. It evaluates models by comparing the relational facts discovered from the test articles with those in Freebase and provides an approximate measure of precision without human evaluation. For evaluation, we draw precision-recall curves for all models. To further verify the effect of our model for  long-tails, we follow  previous studies \cite{han2018hierarchical} to report the Precision@N results. The dataset and baseline code can be found on Github \footnote{https://github.com/thunlp/OpenNRE}.

\subsection{Parameter Settings\footnote{Details of hyper-parameters settings and evaluation  of different instances can be found in  supplementary materials}}
To fairly compare the results of our  models with those baselines, we also set most of the experimental parameters by following \cite{lin2016neural}. We apply dropout on the output layers of our models to prevent overfitting. We also pretrain the sentence encoder of PCNN before training our model.
\begin{figure}
\centering
\includegraphics[width=0.38\textwidth]{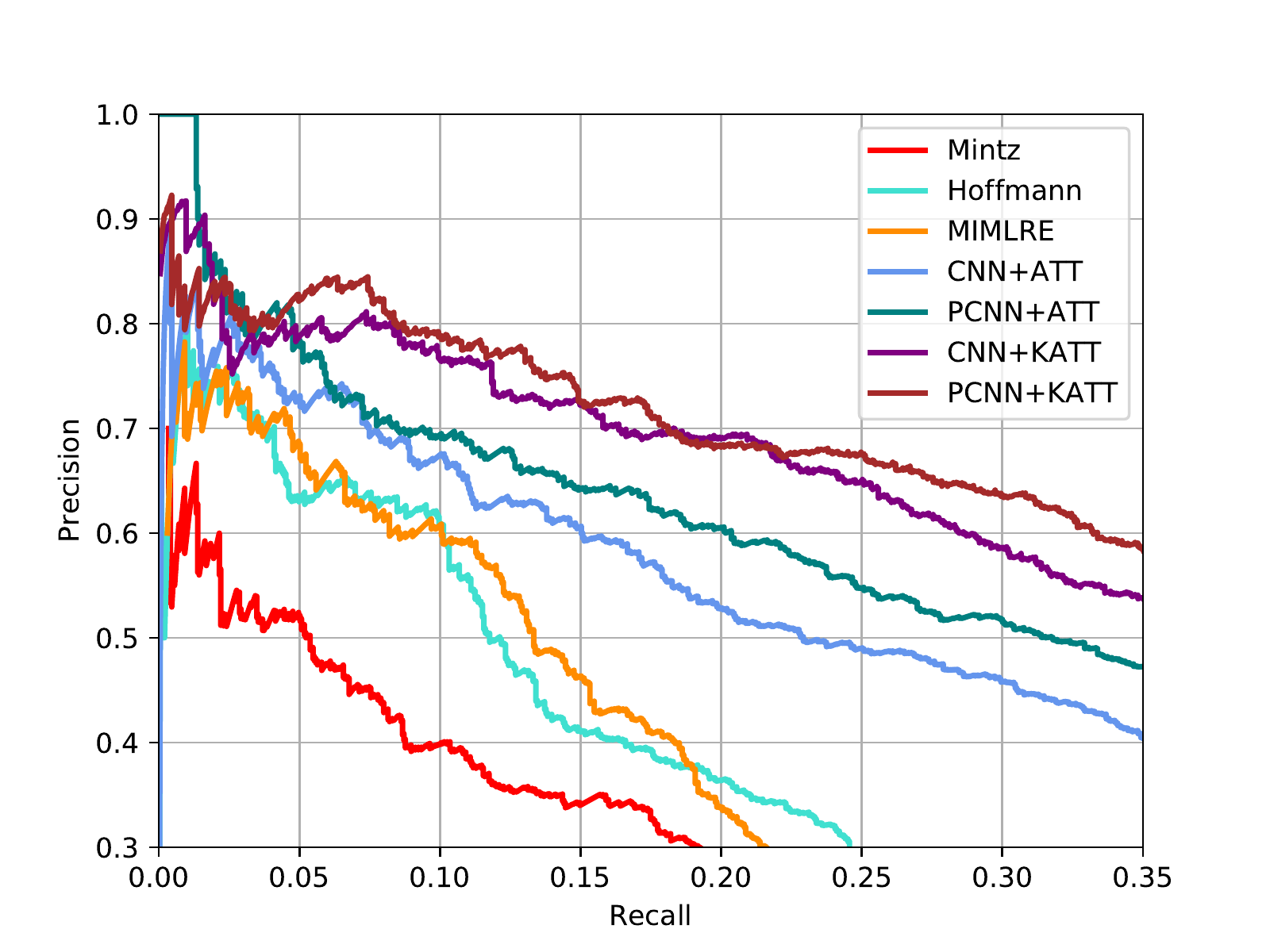}
\caption{\label{pr1}Precision-recall curves for the proposed model and various baseline models.}
\end{figure}
\begin{figure}
\centering
\includegraphics[width=0.38\textwidth]{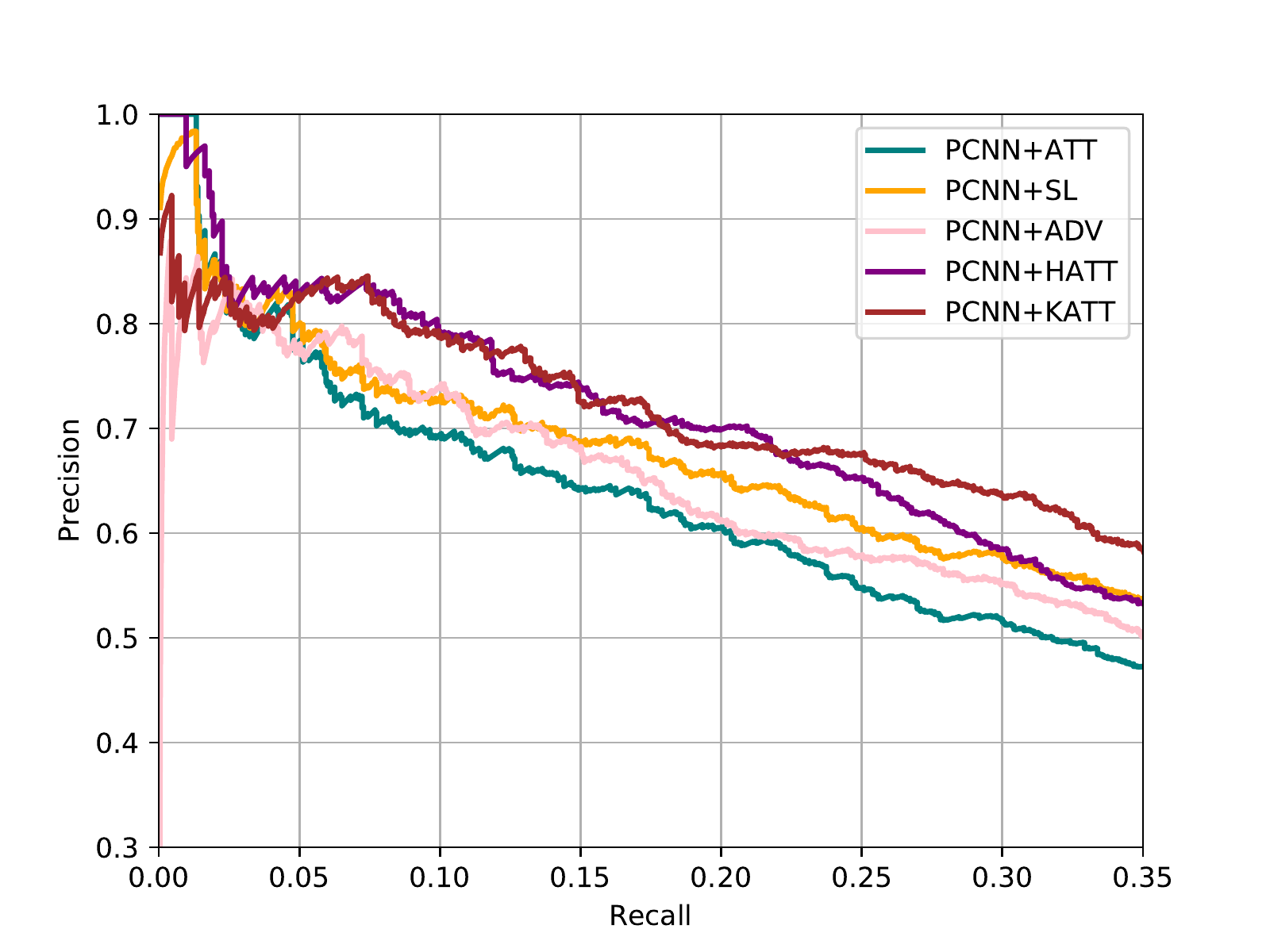}
\caption{\label{pr2}Precision-recall curves for the proposed model and various attention-based neural models.}
\end{figure}
\subsection{Overall Evaluation Results}

To evaluate the performance of our proposed  model, we compare the precision-recall curves of our model with various previous RE models. The evaluation results are shown in Figure \ref{pr1} and Figure \ref{pr2}. We report the results of  neural architectures including CNN and PCNN with various attention based methods: \textbf{+KATT} denotes our  approach, \textbf{+HATT} is  the hierarchical attention  method \cite{han2018hierarchical}, \textbf{+ATT} is the plain selective attention method over instances \cite{lin2016neural}, \textbf{+ATT+ADV} is the denoising attention method by adding a small adversarial perturbation to instance embeddings \cite{wu2017adversarial}, and \textbf{+ATT+SL} is the attention-based model using soft-labeling method to mitigate the side effect of the wrong labeling problem at entity-pair level \cite{liu2017soft}. We also compare our method with feature-based models, including \textbf{Mintz} \cite{mintz2009distant}, \textbf{MultiR} \cite{hoffmann2011knowledge} and \textbf{MIML} \cite{surdeanu2012multi}.

As shown in both  figures, our approach  achieves the best results among all  attention-based models. Even when compared with PCNN+HATT, PCNN+ATT+ADV, and PCNN+ATT+SL, which adopt sophisticated denoising schemes and extra information, our model is still more  advantageous. This indicates that our method can take advantage of the rich correlations between relations through KGs and GCNs, which  improve the performance.  We believe the performance of our model can be further improved by adopting additional mechanisms like adversarial training, and reinforcement learning, which will be part of our future work.

\subsection{Evaluation Results for Long-tail Relations}

\begin{table}[]
    \small
    \begin{tabular}{ccp{0.4cm}p{0.4cm}p{0.4cm}p{0.4cm}p{0.4cm}p{0.4cm}} 
 
        \hline
                \multicolumn{2}{c}{\multirow{2}*{\shortstack{Training Instances \\ Hits@K (Macro)}}} &\multicolumn{3}{c}{$<$100}  &\multicolumn{3}{c}{$<$200} \\
           
              & & 10&15&20&10&15&20 \\
              \hline
                \multirow{3}*{CNN}&+ATT & $<$5.0& $<$5.0 &18.5 &  $<$5.0 &16.2&33.3    \\ 
        
        \cline{3-8}&+HATT&5.6 &31.5 & 57.4 &22.7 & 43.9  & 65.1    \\ 
        
       \cline{3-8}&+KATT&\textbf{9.1} &\textbf{41.3}&\textbf{58.5}&\textbf{23.3}&\textbf{44.1}& \textbf{65.4}   \\ 
            \hline
                \multirow{3}*{PCNN}&+ATT &$<$5.0 & 7.4 & 40.7& 17.2  &24.2&51.5    \\ 
        
        \cline{3-8}&+HATT&29.6 &51.9 & 61.1 &41.4 & 60.6  & 68.2    \\ 
        
       \cline{3-8}&+KATT& \textbf{35.3} &\textbf{62.4}&\textbf{65.1}&\textbf{43.2}&\textbf{61.3}& \textbf{69.2}  \\  \hline
    \end{tabular} 
    \caption{Accuracy (\%) of Hits@K on relations with training instances fewer than 100/200.}
    \label{table2}
\end{table}

\begin{table}[]
    \small 
    \begin{tabular}{cp{0.4cm}p{0.4cm}p{0.4cm}p{0.4cm}p{0.4cm}p{0.4cm}} 
 
    \hline
             \multirow{2}*{\shortstack{Training Instances \\ Hits@K (Macro)}}     
             &\multicolumn{3}{c}{$<$100}  &\multicolumn{3}{c}{$<$200} \\
               &10&15&20&10&15&20 \\
               \hline
   +KATT& \textbf{35.3}&\textbf{62.4}&\textbf{65.1}&\textbf{43.2}&\textbf{61.3}& \textbf{69.2}  \\ 
   \hline
      w/o hier&34.2 &62.1  & 65.1& 42.5&60.2 & 68.1  \\ 
   
    w/o GCNs&30.5 &61.9  &63.1&39.5&58.4& 66.1  \\ 
  
    Word2vec&30.2 &62.0  &62.5&39.6&57.5& 65.8   \\ 
  
     w/o KG& 30.0 &61.0  &61.3&39.5&56.5& 62.5  \\  
   \hline
   
    \end{tabular} 
    \caption{Results of ablation study with PCNN.}
    \label{table4}
\end{table}

To further demonstrate the improvements in performance for long-tail relations, following the study by \cite{han2018hierarchical} we extract a subset of the test dataset in which all the relations have fewer than 100/200 training instances. We employ the Hits@K metric for evaluation. For each entity pair, the evaluation requires its corresponding golden relation in the first $K$ candidate relations recommended by the models. Because it is difficult for the existing models to extract long-tail relations, we select $K$ from \{10,15,20\}. We report the  macro average Hits@K accuracies for these subsets because the micro-average score generally overlooks the influences of  long-tails. From the results shown in Table \ref{table2}, we observe that  for both CNN and PCNN models, our  model outperforms the plain attention model and the HATT model.  Although our KATT method has achieved better results for long-tail relations as compared to both   plain ATT method and HATT method, the results of all these methods are still far from satisfactory. This indicates that distantly supervised RE models  still suffer from the long-tail relation problem,  which may require  additional schemes and extra information to solve this problem in the future.

\begin{figure*} \centering
\subfigure[HATT] { \label{aa}
  \includegraphics[width=0.23\textwidth]{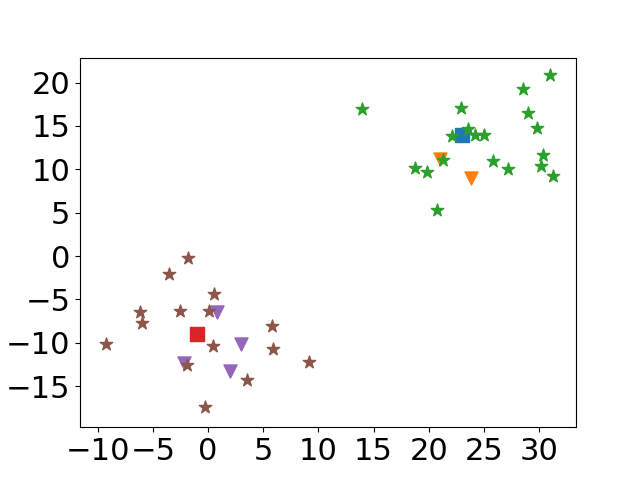}
}
\subfigure[+KG ] { \label{cc}
\includegraphics[width=0.23\textwidth]{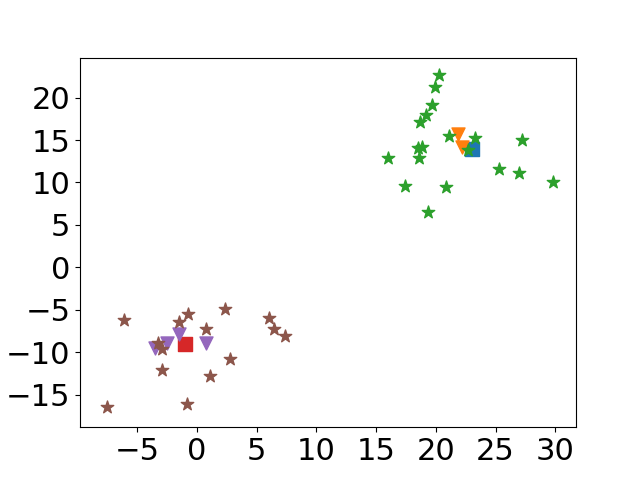}
}
\subfigure[+GCNs] { \label{bb}
\includegraphics[width=0.23\textwidth]{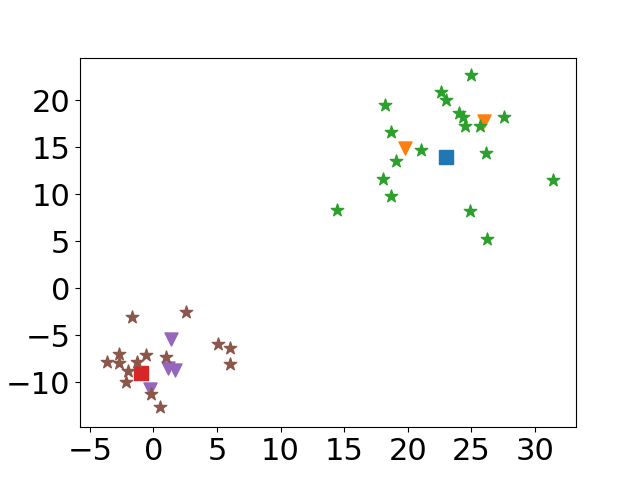}
}
\subfigure[KATT] { \label{dd}
\includegraphics[width=0.23\textwidth]{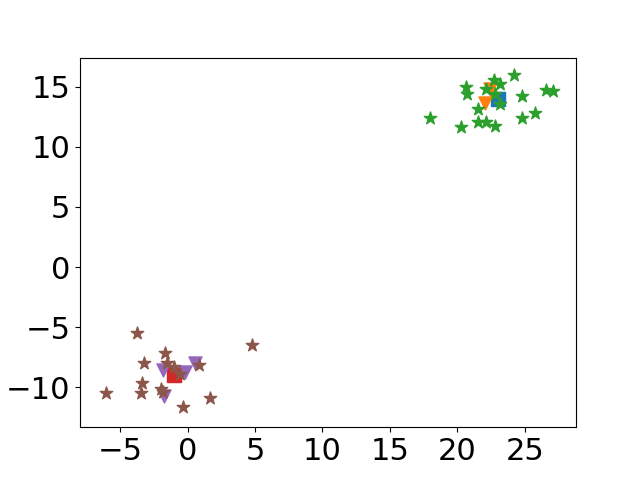}
}
\caption{T-SNE visualizations of  class embeddings. Cluster in the upper right  is the relation /location/*/* and cluster in the bottom left  is the  relation /people/*/* ). The  square,  triangle, and star   refer to the high (/location)  middle  (/location/location/) and  base (/location/location/contains) level relations,  respectively. }
\label{4pic}
\end{figure*}

\subsection{Ablation Study}
To analyze the contributions and effects of different technologies in our approach,  we perform ablation tests.  \textbf{+KATT} is our  method; \textbf{w/o hier} is the method without coarse-to-fine attention (only utilizes  bottom node  embeddings of the hierarchy label graph), which implies no knowledge transfer from its higher level classes; \textbf{w/o GCN} is the method without GCNs, which implies no explicit relational knowledge; \textbf{Word2vec} is the method in which the node is initialized  with pretrained Skip-Gram \cite{mikolov2013efficient} embeddings; and \textbf{w/o KG} is the method in which the node is  initialized with random embeddings, which implies  no prior  relational knowledge from KGs. From the evaluation results in Table \ref{table4}, we observe that the performance slightly degraded without  coarse-to-fine attention, which proves that knowledge transfer from the higher node is useful. We also  noticed that the performance slightly degraded without KG or using word embeddings, and the performance significantly degraded when we removed GCNs. This is reasonable because  GCNs can learn  more explicit correlations between relation labels, which boost the performance for long-tail relations.

\begin{table}[h]\centering
    \fontsize{8.5}{10}\selectfont
    \begin{tabular}{|p{4.9cm}|p{0.75cm}|p{0.75cm}|}
    \hline 
    /people/deceased\_person/place\_of\_burial&\textbf{HATT}&\textbf{KATT} \\
    \hline
    \textbf{richard\_wagner} had his \textbf{bayreuth}, with its festspielhaus specially designed to accommodate his music dramas. &0.21  & 0.07 \\
        \hline
wotan and alberich are \textbf{richard\_wagner}; and the rheingold and valhalla are wagner's real-life grail, the opera house in \textbf{bayreuth}.& 0.15& 0.13  
    \\
    \hline
    /location/br\_state/capital & \textbf{HATT} &\textbf{KATT} \\
       
     \hline
    there's violence everywhere, said ms. mesquita, who, like her friend, lives in \textbf{belo\_horizonte}, the capital of \textbf{minas\_gerais} state&0.47 &0.51\\
     \hline all the research indicates that we are certain to find more gas in th amazon, eduardo braga, the governor of \textbf{amazonas}, said in an interview in \textbf{manaus}&  0.46&0.45\\
   \hline
   
    \end{tabular}
    \caption{Example sentences for case study.}
    \label{table5}
\end{table}

\subsection{Case Study}
We give some examples to show  how our method  affects the selection of  sentences. In Table \ref{table5}, we display the sentence's attention score in the lowest level\footnote{Both  HATT and KATT methods can successfully select  the correct sentence at the higher-level;  details can be found in supplementary materials.}.  
Both the relation /people/deceased\_person/place\_of\_burial (24 instances) and  /location/br\_state/capital  (4 instances) are   long-tail relations. On one hand, relation /people/deceased\_person/place\_of\_burial  has semantically similar data-rich relation such as /people/deceased\_person/place\_of\_death.  
We observe  that  HATT erroneously assigns high attention to the incorrect sentence whereas KATT  successfully  assigns the right attention weights, which demonstrates the efficacy  of  knowledge  transfer from semantically similar relations (Both  HATT and KATT methods can take advantage of knowledge transfer of high-level relations.).  On the other hand, the relation /location/br\_state/capital does not have semantically similar relations. However, we notice that KATT  still successfully assigns the right attention weights,  which demonstrates the efficacy  of  knowledge  transfer from high-level relations   using coarse-to-fine knowledge-aware attention.  

\subsection{Visualizations of Class Embeddings}

We  visualize  the class embeddings via t-SNE \cite{maaten2008visualizing}    to  further show how GCNs and KG embeddings can help RE for long-tail relations.   We observe  that (1) Figure  \ref{aa}  and  \ref{dd}   show that  semantically  similar class embeddings are closer with GCNs and pretrained KG embeddings,  which help select  long-tail instances. (2) Figure    \ref{cc}  and \ref{bb} show that  KG embeddings  and GCNs  have different  contributions for different relations to learn   relational knowledge between classes. For example, /location/location/contain has a sparse hierarchy structure, which leads to inefficient learning for GCNs; therefore,  the relative distance changes only slightly, which reveals the necessity of implicit  relational knowledge from KGs. (3) Figure  \ref{dd} shows that there are still some semantically similar class embeddings located far away, which may degrade the performance  for long-tails. This may be caused by  either sparsity in the hierarchy label graph  or  equal  treatment for nodes with the same parent  in  GCNs, which is not a reasonable hypothesis.  We will address this by integrating more  information  such as relation descriptions  or combing logic reasoning as a part of future work.

\section{Conclusion and Future Work}
In this paper, we take advantage of the knowledge from  data-rich classes at the head of distribution to boost the performance of the data-poor classes at the tail.  As compared to previous  methods, our approach provides fine-grained  relational  knowledge  among classes  using   KG  and GCNs, which is quite effective and  encoder-agnostic.

In the future, we plan to explore the following directions: (1) We may  combine our  method with recent denoising methods to further improve  performance. (2) We may  combine  rule mining and reasoning technologies  to learn better   class embeddings to boost performance.  (3) It will be promising to apply our method to zero-shot RE and further adapt to other NLP scenarios.
 
\section*{Acknowledgments}
We  want to express gratitude to the anonymous reviewers for their hard work and kind comments, which will further improve our work in the future. This work is  funded by NSFC91846204/61473260, national key research program YS2018YFB140004,  Alibaba CangJingGe (Knowledge Engine) Research Plan and Natural  Science Foundation of Zhejiang Province of China (LQ19F030001).


\bibliography{sample}

\bibliographystyle{acl_natbib}

\end{document}


\maketitle

\section{Hierarchy Label Graph Construction}
Previous study \cite{xie2017interpretable,zhang2018knowledge} indicate that there is rich information from the relation structure.   Some dataset has explicit  relation hierarchical structures. For example, the relation /location/province/capital in Freebase \cite{bollacker2008freebase} indicates the relation between a province and its capital. It is labeled under the location branch. Under this branch, there are some other relations /location/location/contains and /location/country/capital, which is closely correlated to each other. However, some other datasets may not have predefined hierarchical structures. It is intuitive to define a  hierarchical relation structure, which can be confirmed by relation clusters and relations in KGs \cite{zhang2018knowledge}.  We utilize three simple methods to build  relation structure and tests their results. All pretrained knowledge graph embeddings are available online\footnote{http://openke.thunlp.org/}. 

\textbf{K-means.} We  first run TransE \cite{bordes2013translating} on a given data set and obtain the embeddings of relations $r_1, r_2, r_3, ..., r_{|R|}$
, where $|R|$ is the number of relations following \cite{zhang2018knowledge}. Then, the k-means algorithm is applied to these embeddings. In this way, we get relation clusters $C_1, C_2, C_3, ..., C_{|C|}$, where $C$ is the set of relation clusters. Previous studies have shown that the embeddings of semantically similar relations locate near each other in the latent space \cite{yang2014embedding}. In this way, we can find relation clusters composed of semantically related relations. Practically, we can get relation clusters recursively and get the hierarchy of relations.

\textbf{Hierarchical Clustering (HC).} Hierarchical clustering  is a method of cluster analysis which seeks to build a hierarchy of clusters. We adopt the  agglomerative way which is a "bottom-up" approach to construct the hierarchy of classes.  Each class starts in its cluster, and pairs of clusters are merged as one moves up the hierarchy. Specifically, we apply Euclidean distance as appropriate metric between classes  and  complete-linkage clustering to  take the distance between clusters.

\textbf{Association Rule Mining under Incomplete Evidence (AMIE).}
AMIE is a  rule mining model that is explicitly tailored to
support the open world assumption scenario. It is inspired by association rule
mining and uses a novel measure for confidence. 
Practically, we run the AMIE tool on the dataset and choose those classes with pca confidence greater than the threshold as correlated classes and then group them. AMIE tool is available oneline\footnote{http://resources.mpi-inf.mpg.de/yago-naga/amie/downloads.html}.

We evaluate the performance of our proposed  model with different hierarchical classes structures on the same dataset NYT. We employ the Hits@K metric for evaluation. For each entity pair, the evaluation requires its corresponding golden relation in the first $K$ candidate relations recommended by the models. Because it is difficult for the existing models to extract long-tail relations, we select $K$ from \{10,15,20\}. We report the  macro average Hits@K accuracies for these subsets since the micro-average score generally overlooks the influences of the long-tails. We report the results of variations of our methods: \textbf{+KATT (FB)} is our  approach with predefined hierarchies; \textbf{K-means (3)} and \textbf{K-means (4)}  are K-means clustering methods to construct class  hierarchies with three and four layers
respectively; \textbf{HC (3)} and \textbf{HC (4)} are hierarchical clustering  methods to construct class hierarchies with three and four layers respectively; \textbf{AMIE} is the association rule mining under incomplete evidence to construct class hierarchies.

From the results in Table \ref{table4},  we observe: (i)
our  model outperforms all the variations, which indicates that data-driven clustering may still have noise and may construct a wrong hierarchy between classes to degrade the performance;  
(ii)  data-driven clustering methods still outperform  vanilla PCNN  which further proves  that  relational knowledge  between labels provide a much informal reference  which boosts the performance for long-tail relations;  
(iii)  K-means (4)  outperforms  K-means (3)   and HC (4)  outperforms HC (3)  which indicate the contributions of complex structures; (vi) K-means methods achieve comparable  performance with AMIE and outperforms  HC. This  may be caused by the  complexity and ambiguity of language as simple L2 distance may not inadequate to measure the similarity which leads to the wrong clustering.

\section{Initialization and Implementation Details}
To fairly compare the results of our approach with those baselines, we set most of the experimental parameters following \cite{lin2016neural}. Table \ref{nyt} shows all experimental parameters used in the experiments. We apply dropout on the output layers of our models to prevent overfitting. For CNN, we set the dropout rate to 0.5. For PCNN, we set the dropout rate to 0.9 to alleviate the overfitting problem. We also pre-train the sentence encoder of PCNN before training our hierarchical attention. 

\begin{table}[]
    \small 
    \begin{tabular}{cp{0.4cm}p{0.4cm}p{0.4cm}p{0.4cm}p{0.4cm}p{0.4cm}} 
 
    \hline
             \multirow{2}*{\shortstack{Training Instances \\ Hits@K (Macro)}}     
             &\multicolumn{3}{c}{$<$100}  &\multicolumn{3}{c}{$<$200} \\
               &10&15&20&10&15&20 \\
               \hline
   +KATT (FB)& \textbf{35.3}&\textbf{62.4}&\textbf{65.1}&\textbf{43.2}&\textbf{61.3}& \textbf{69.2}  \\ 
   \hline
      K-means (3)&34.2 &62.1  & 64.1& 42.5&61.3 & 68.1  \\ 
      K-means (4)&34.3 &62.3 & 64.3& 42.7&61.4 & 68.1  \\ 
           HC (3)&30.2 &62.1  & 60.1& 40.5&60.2 & 66.1  \\ 
           HC (4)&31.2 &61.1  & 62.1& 41.5&60.2 & 66.2  \\ 

  AMIE&34.2 &62.1  & 62.3& 42.2&61.2 & 67.9  \\ 
  \hline
   PCNN &$<$5.0 & 7.4 & 40.7& 17.2  &24.2&51.5     \\ 
   \hline
    \end{tabular} 
    \caption{Evaluation of our approach with different hierarchy label graph construction methods for relations with training instances fewer than 100/200.}
    \label{table4}
\end{table}

\begin{table}[!htbp]
\centering

\begin{tabular}{c|c}

\hline
 \centering \textbf{Parameter}&  \textbf{Settings}\\
\hline
 \centering Convolution Kernel Size& 3 \\
 \centering Word Embedding Size&300  \\
\centering Position Embedding Size& 5 \\
\centering Learning Rate& 0.1 \\
\centering Hidden Layer Dimension  for CNNs&230 \\
\centering Batch Size&160  \\

\hline
\end{tabular}\caption{Parameter settings.}\label{nyt}

\end{table}
\begin{table}[h]\centering
    \fontsize{8.5}{10}\selectfont
    \begin{tabular}{|p{4.9cm}|p{0.75cm}|p{0.75cm}|}
    \hline 
    /people/deceased\_person/place\_of\_burial&\textbf{HATT}&\textbf{KATT} \\
    \hline
    \textbf{richard\_wagner} had his \textbf{bayreuth}, with its festspielhaus specially designed to accommodate his music dramas. &0.03  & 0.07 \\
        \hline
wotan and alberich are \textbf{richard\_wagner}; and the rheingold and valhalla are wagner's real-life grail, the opera house in \textbf{bayreuth}.& 0.15& 0.13  
    \\
    \hline
    /location/br\_state/capital & \textbf{HATT} &\textbf{KATT} \\
       
     \hline
    there's violence everywhere, said ms. mesquita, who, like her friend, lives in \textbf{belo\_horizonte}, the capital of \textbf{minas\_gerais} state&0.48 &0.53\\
     \hline all the research indicates that we are certain to find more gas in th amazon, eduardo braga, the governor of \textbf{amazonas}, said in an interview in \textbf{manaus}&  0.56&0.55\\
   \hline
   
    \end{tabular}
    \caption{Example sentences for case study in higher-level.}
    \label{table5}
\end{table}

In order to relieve the impact of $NA$ in DS based
relation extraction, we cut the propagation of loss from $NA$ following \cite{ye2017jointly}, which means if relation $c$ is $NA$, we set its loss as $0$. 

\section{Case Study}

We report  some additional  examples of how our method   takes effect in selecting the sentences in the higher level. In Table \ref{table5}, we display the sentence's attention score in the higher level (e.g., attention weights of node /people/deceased\_person/).  
Both the relation /people/deceased\_person/place\_of\_burial (24 instances) and  /location/br\_state/capital  (4 instances) are   long-tail relations. For those relations, the instance recommended by the higher-level attention straightforwardly expresses the relational facts.
This example shows that our approach achieves comparable results with HATT \cite{han2018hierarchical} and is beneficial for these long-tail relations .

\section{Evaluation Results for Different Instances}

 \begin{table*} \centering
   \small
    \begin{tabular}{ccccccccccccc} 
    \hline
    Test Mode &\multicolumn{4}{c}{ONE} &\multicolumn{4}{c}{TWO}& \multicolumn{4}{c}{ALL} \\
    \hline
    P@N&100&200&300&Mean&100&200&300&Mean&100&200&300&Mean \\
    \hline
   CNN+ONE&68.3&60.7&53.8&60.9&70.3&62.7&55.8&62.9&67.3&64.7&58.1&63.4\\
   CNN+ATT&76.2&65.2&60.8&67.4&76.2&65.7&62.1&68.0&76.2&68.6&59.8&68.2\\
  CNN+HATT&88.0&74.5&68.0&76.8&85.0&76.0&73.0&78.0&88.0&79.0&77.7&81.6\\
  CNN+KATT&\textbf{90.0}&\textbf{75.1}&\textbf{69.3}&\textbf{79.0}&\textbf{85.3}&\textbf{79.2}&\textbf{74.0}&\textbf{79.5}&\textbf{89.2}&\textbf{80.2}&\textbf{79.2}&\textbf{82.9}\\
    \hline
    PCNN+ONE&73.3&64.8&56.8&65.0&70.3&67.27&63.1&66.9&72.3&69.7&64.1&68,7\\
    PCNN+ATT&73.3&69.2&60.8&67.8&77.2&71.6&66.1&71.6&76.2&73.1&67.4&72.2\\
   PCNN+HATT&84.0&76.0&69.7&76.6&85.0&76.0&72.7&77.9&88.0&79.5&75.3&80.9\\
   PCNN+KATT&\textbf{85.3}&\textbf{77.1}&\textbf{70.2}&\textbf{77.5}&\textbf{86.1}&\textbf{77.1}&\textbf{73.1}&\textbf{78.8}&\textbf{89.2}&\textbf{80.1}&\textbf{75.8}&\textbf{81.7}\\
                \hline
    \end{tabular}
    \caption{Top-N precision (P@N) for RE on the entity pairs with different number of instances (\%).}
    \label{table3}
\end{table*}

Since our method  focuses on modifications over selective attention, we also conduct Precision@N tests on those entity pairs with few instances following \cite{lin2016neural,han2018hierarchical}. We use the three test settings for this experiment: the ONE test set where we randomly select one instance for each entity pair for evaluation; the TWO test set where we randomly select two instances for each entity pair; the ALL test set where we use all instances for each remaining entity pair for evaluation. For the ONE and TWO test set, we intend to show that taking correlation information among relations into consideration can lead to a better relation
classifier. The ALL test set is designed to show the effect of our attention over multiple instances. We report the precision values of top $N$ triples extracted, where $N \in \{100,200,300\}$.

The evaluation results are shown in Table \ref{table3}, and from the results, we observe that (1) The performance of all methods is generally improved as the instance number increases which shows that the selective attention model can effectively take advantage of information from multiple noisy instances by combining useful instances while discarding useless ones. (2) Our KATT method has higher precision values in the ONE test set which indicates that  in a few-shot scenario, correlations  among relations can be caught by our model.

\bibliography{sample}

\bibliographystyle{acl_natbib}